\documentclass[aps,prl,showpacs]{revtex4}
\usepackage{graphicx}
\usepackage{amsfonts,amssymb,amsmath}
\usepackage{bm}

\begin{document}

\title{Solution of the Schr{\"o}dinger equation with one and two dimensional double-well potentials}
\date{\today}
 
\author{Ramazan Ko\c{c}}
\email{koc@gantep.edu.tr}
\affiliation{Department of Physics, Faculty of Engineering 
University of Gaziantep,  27310 Gaziantep, Turkey}
\author{Derya Haydargil}
\email{haydargil@gantep.edu.tr}
\affiliation{Department of Physics, Faculty of Engineering 
University of Gaziantep,  27310 Gaziantep, Turkey}

\begin{abstract}
 The Schr\"{o}dinger equation with one and two dimensional potentials are
 solved in the frame work of the $sl_{2}(R)$ Lie algebra. Eigenfunctions of
 the Schr\"{o}dinger equation for various asymmetric double-well potentials
 have been determined and the eigenstates are expressed in terms of the
 orthogonal polynomials. The solution of the double-well potential in two
 dimension have been analyzed.

\end{abstract}

\maketitle

\section{Introduction}

The importance of the double-well potentials in quantum mechanics, condensed
matter physics, statistical physics or field theory can hardly be
overestimated \cite{benderski, baroni}. The Schr\"{o}dinger equation has no
exact, analytical solution with double-well potentials. It can be solved by
using approximate methods or numerical methods. On the other hand, in
quantum mechanics there exist potentials for which it is possible to find a
number of eigenvalues and associated eigenfunctions exactly and in a closed
form. These systems are said to be quasi-exactly solvable (QES) and this
property is ultimately connected with the existence of a hidden dynamical
algebra \cite{turbiner1}.

In this article we present a quasi exact solution of the Schr\"{o}dinger
equation with one- and two dimensional double-well potentials. The
Schr\"{o}dinger equation with one dimensional double-well potential have
been studied by a number of authors and it can be solved by using various
methods \cite{konwent}. Approximate QES and approximate analytical solutions
of the double-well potential have been derived in \cite{burrow}.

There is not much attention for the solution of the two-dimensional
double-well potential \cite{zaslavski}. In this paper we develop a simple
method to solve the Schr\"{o}dinger equation which admits separation of
variables with two-dimensional double-well potentials. We show the separated
equations are still QES. The Schr\"{o}dinger equation does not admit
separation of variables in two dimensions, a topic of a future work.

The paper is organized as follows. Section 2 is devoted to a derivation of a
QES one-dimensional double-well potential in the framework of the $sl_{2}(R)$
Lie algebra. Section 3 contains a solution of the two dimensional double-well
potential which admits separation of variables. Finally, the results are
discussed in section 4.

\section{One dimensional double-well potential}

The strategy we follow is described in \cite{turbiner2}. One way to
construct a quasi-exactly solvable differential equation is to express the
Hamiltonian in terms of the generators of the Lie algebra. Let us consider
the $sl_{2}(R)$ algebra realized as follows:

\begin{equation}
J_{+}=x^{2}\frac{d}{dx}-jx,\qquad J_{0}=x\frac{d}{dx}-\frac{j}{2},\qquad
J_{-}=\frac{d}{dx}.  \label{eq:1}
\end{equation}
The generators obey the commutation relation
\begin{equation}
\lbrack J_{+},\,J_{-}]=-2J_{0},\qquad \lbrack J_{0},\,J_{\pm }]=\pm \,J_{\pm
},  \label{eq:2}
\end{equation}
and it may be verified that the eigensolutions of the generators are given
explicitly by
\begin{equation}
J_{-}|\,m>=m\,|\,m-1>,\quad J_{0}\,|\,m>=\left( m-\frac{j}{2}\right)
|\,m\!>,\quad J_{+}\,|\,m\!>=(m-j)\,|\,m+1),  \label{eq:3}
\end{equation}
where the basis function is
\begin{equation}
|\,m>=R_{j+1}(x).  \label{eq:4}
\end{equation}
If $j$ is a non-negative integer number, the algebra possesses a $(j+1)$%
-dimensional subspace
\begin{equation}
R_{j+1}=\left\langle \,x^{0},x^{2},\cdots ,x^{j}\,\right\rangle .
\label{eq:5}
\end{equation}
The linear and bilinear combinations of the generators of the algebra form a
second order differential equation possessing polynomial solutions. One
possible way to get a QES Schr\"{o}dinger operator is to transform the QES
algebraic operators acting on the finite dimensional subspace of polynomials
R$_{j+1}(x)$ into the Schr\"{o}dinger-type equation. It can be done by a
gauge transformation.

\subsection{\protect\bigskip The model Potential I}

Consider the double-well potential of the form
\begin{equation}
V(x)=\frac{A}{2}x^{2}-\frac{B}{3}x^{3}+\frac{C}{4}x^{4},  \label{eq:6}
\end{equation}
where $A,$ $B$ and $C$ are constants. If $2B^{2}=9AC$, the potential is a
symmetric potential with two minima; otherwise the potential is an
asymmetric potential with two minima. We are interested in the asymmetric
double-well potential. The Schr\"{o}dinger equation of this potential can be
written as
\begin{equation}
\left( -\frac{d^{2}}{dx^{2}}+V(x)\right) \psi (x)=E\psi (x).  \label{eq:7}
\end{equation}
We can write the wavefunction $\psi (x)$ in the form
\begin{equation}
    \psi (x)=\exp \left( -\frac{1}{2}\left( -3(j+1)\frac{Cx}{B}-\frac{Bx^{2}}{3%
    \sqrt{C}}+\frac{\sqrt{C}x^{3}}{3}\right) \right) R(x).  \label{eq:8}
\end{equation}
Then the Schr\"{o}dinger equation takes the form
\begin{eqnarray}
    &&-\frac{d^{2}R(x)}{dx^{2}}+\left( -3(j+1)\frac{C}{B}-\frac{2B}{3\sqrt{C}}x+%
    \sqrt{C}x^{2}\right) \frac{dR(x)}{dx}-  \notag \\
    &&\left( \varepsilon +j\left( -\frac{B}{3\sqrt{C}}+\sqrt{C}x\right) \right)
    R(x)=0  \label{eq:9}
\end{eqnarray}
with the constraint
\begin{eqnarray}
    A &=&\frac{\left( 2B^{3}-27(j+1)C^{5/2}\right) }{9BC}  \notag \\
    E &=&\varepsilon -\frac{(j+1)B}{3\sqrt{C}}-\left( \frac{3(j+1)C}{2B}\right)
    ^{2}.  \label{eq:10}
\end{eqnarray}
The potential (\ref{eq:6}) is asymmetric for all values of $j$ when $B>0$
and $C>0$, and the location of minimum points depends on the values of $j$.
In order to solve (\ref{eq:9}), we introduce the linear and
bilinear combination of the algebraic operators
\begin{equation}
T=-J_{-}^{2}-3(j+1)\frac{C}{B}J_{-}-\frac{2B}{3\sqrt{C}}J_{0}+\sqrt{C}J_{+}
\label{eq:11}
\end{equation}
for which one can define the spectral problem
\begin{equation}
\text{TR}(x)=\varepsilon R(x),  \label{eq:12}
\end{equation}
where $\varepsilon $ is a spectral parameter. The algebraic structure (\ref
{eq:11}) is quasi-exactly solvable and the insertion of (\ref{eq:1}) into (%
\ref{eq:11}) leads to the differential equation (\ref{eq:9}). The basis
function (\ref{eq:5}) can be written as
\begin{equation}
R(x)=\sum_{m=0}^{\infty }a_{m}P_{m}(\varepsilon )x^{m},  \label{eq:13}
\end{equation}
where $P_{m}(\varepsilon )$ is polynomial in $\varepsilon $ and it can be
given by the recurrence relation
\begin{eqnarray}
&&-m(m-1)P_{m-2}(\varepsilon )-3m(j+1)\frac{C}{B}P_{m-1}(\varepsilon )+\sqrt{%
C}\left( m-j\right) P_{m+1}(\varepsilon )-  \notag \\
&&\left( \frac{2B}{3\sqrt{C}}(m-\frac{j}{2})+\varepsilon \right)
P_{m}(\varepsilon )=0,  \label{eq:14}
\end{eqnarray}
with the initial condition $P_{0}=1.$ The spectral parameter $\varepsilon $
can be determined from the recurrence relation (\ref{eq:14}). If $%
\varepsilon _{j}$ is a root of the polynomial $P_{m+1}(\varepsilon )$, the
series (\ref{eq:13}) truncates at $m\eqslantgtr (j+1)$ and $\varepsilon _{j}$
belongs to the spectrum of the double-well potential. The first four of
these polynomials are given by
\begin{eqnarray}
P_{1}(\varepsilon ) &=&\varepsilon   \notag \\
P_{2}(\varepsilon ) &=&9BC\varepsilon ^{2}-B^{3}-54C^{5/2}  \notag \\
P_{3}(\varepsilon ) &=&9BC\varepsilon ^{3}-4\left( B^{3}+81C^{5/2}\right)
\varepsilon +36BC^{2}  \notag \\
P_{4}(\varepsilon ) &=&9B^{2}C^{2}\varepsilon ^{4}-\left(
10B^{4}C+1080BC^{7/2}\right) \varepsilon ^{2}+  \notag \\
&&216B^{2}C^{3}\varepsilon +\left( B^{6}+216B^{3}C^{5/2}+11664C^{5}\right)
\label{eq:15}
\end{eqnarray}
for $j=0,1,2$ and $3$, respectively. Analytical solutions of the recurrence
relation (\ref{eq:14}) are available only for the first few values of $j$.
For $j>2$ the solutions become numerical.

\subsection{The Model Potential II}

We consider the following potential treated by Burrows \cite{burrow}
\begin{equation}
V(x)=V_{0}-Ax+\frac{x^{2}}{2}(B+Cx)^{2}.  \label{eq:16}
\end{equation}
The potential is symmetric when $A=0$, with double minima at $x=-\frac{B}{C}$
and $x=-\frac{B}{2C}$. The Schr\"{o}dinger equation with the potential (\ref
{eq:16}) is given by
\begin{equation}
\left( -\frac{d^{2}}{dx^{2}}+V(x)\right) \psi (x)=E\psi (x).  \label{eq:17}
\end{equation}
In order to transform the Schr\"{o}dinger equation to the form of an
algebraic equation, we introduce the following transformation:
\begin{equation}
    \psi (x)=\exp \left( -\frac{Bx^{2}}{2\sqrt{2}}-\frac{Cx^{3}}{3\sqrt{2}}%
    \right) R(x).  \label{eq:18}
\end{equation}
Then the Schr\"{o}dinger equation takes the form
\begin{eqnarray}
    &&-\frac{d^{2}R(x)}{dx^{2}}-\sqrt{2}x(B+Cx)\frac{dR(x)}{dx}+  \label{eq:19}
    \\
    &&\left( V_{0}-E-Ax+\sqrt{2}Cx-\frac{B}{\sqrt{2}}\right) R(x)=0.  \notag
\end{eqnarray}
The linear and bilinear combinations of the operators of the Lie algebra
\begin{equation}
    T=-J_{-}^{2}-\sqrt{2}BJ_{0}-\sqrt{2}CJ_{+};\quad TR(x)=\varepsilon R(x)
    \label{eq:20}
\end{equation}
is equivalent to the situation in (\ref{eq:19}) when the
parameters of the Schr\"{o}dinger equation is constrained to
\begin{eqnarray}
A &=&-\sqrt{2}C(j+1)  \notag \\
E &=&\varepsilon -\frac{(j+1)B}{\sqrt{2}}.  \label{eq:21}
\end{eqnarray}
Since $j\eqslantgtr 0$, the potential is an asymmetric double-well
potential. The Spectral parameter $\varepsilon $ can be calculated from the
recurrence relation
\begin{eqnarray}
&&-m(m-1)P_{m-2}(\varepsilon )-\left( \sqrt{2}B\left( m-\frac{j}{2}\right)
+\varepsilon \right) P_{m}(\varepsilon )-  \label{eq:22} \\
&&\sqrt{2}C(m-j)P_{m+1}(\varepsilon )=0,  \notag
\end{eqnarray}
with the initial condition $P_{0}(\varepsilon )=1.$ The first four values of
the $P_{m}(\varepsilon )$ are given as follows:
\begin{eqnarray}
P_{1}(\varepsilon ) &=&\varepsilon   \notag \\
P_{2}(\varepsilon ) &=&2\varepsilon ^{2}-B^{2}  \notag \\
P_{3}(\varepsilon ) &=&\sqrt{2}\left( \varepsilon ^{3}-2\varepsilon
B^{2}-4BC\right) +4\varepsilon C  \notag \\
P_{4}(\varepsilon ) &=&4\varepsilon ^{4}-20\left( B^{2}-2\sqrt{2}C\right)
\varepsilon ^{2}+96\varepsilon BC+9B^{2}\left( B^{2}-4\sqrt{2}C\right) ,
\label{eq:23}
\end{eqnarray}
for $j=0$, $1$, $2$, $3$, respectively.

\section{Two-Dimensional double-well potential}

In this section the quasi exact solution of the Schr\"{o}dinger equation
with two dimensional double-well potential is treated. We develop a
systematic procedure for constructing a QES of a double-well potential and
we illustrate our method in the situation where the Schr\"{o}dinger equation
admits separation of variables. The Schr\"{o}dinger equation in two
dimensions is given by
\begin{equation}
    \left( -\frac{\partial ^{2}}{\partial x^{2}}-\frac{\partial ^{2}}{\partial
    y^{2}}+V(x,y)\right) \psi (x,y)=E\psi (x,y),  \label{eq:24}
\end{equation}
and can be solved exactly for a few potentials. The ground state wave
function of the Schr\"{o}dinger equation can be written in the form
\begin{equation}
\psi _{0}(x,y)=\exp (-W(x)-F(y)-f(xy)),  \label{eq:25}
\end{equation}
where $W(x)$ and $F(y)$ are functions of $x$ and $y,$ respectively, while $%
f(xy)$ is function of the product of $x$ and $y$. The potential and ground
state energy of the system with the given ground state wave function takes
the form
\begin{equation}
V_{0}(x,y)-E_{0}=\frac{d^{2}W}{dx^{2}}+\frac{d^{2}F}{dy^{2}}+\frac{\partial
^{2}f}{\partial x^{2}}+\frac{\partial ^{2}f}{\partial y^{2}}-\left( \frac{dW%
}{dx}+\frac{\partial f}{\partial x}\right) ^{2}-\left( \frac{dF}{dy}+\frac{%
\partial f}{\partial y}\right) ^{2}.  \label{eq:26}
\end{equation}
When $f(xy)$ is zero, then the equation is separable and the general
solution of the system can be obtained by solving two one dimensional
differential equation. The wave function of the excited states can be
written as
\begin{equation}
\psi (x,y)=\phi (x,y)\psi _{0}(x,y),  \label{eq:27}
\end{equation}
where the function $\phi (x,y)$ is polynomial in $x$ and $y$. Substituting (%
\ref{eq:27}) into Schr\"{o}dinger equation (\ref{eq:24}) we obtain the
following differential equation:
\begin{eqnarray}
&&-\frac{\partial ^{2}\phi }{\partial x^{2}}-\frac{\partial ^{2}\phi }{%
\partial y^{2}}-2\left( \frac{dW}{dx}+\frac{\partial f}{\partial x}\right)
\frac{\partial \phi }{\partial x}-2\left( \frac{dF}{dy}+\frac{\partial f}{%
\partial y}\right) \frac{\partial \phi }{\partial y}+  \label{eq:28} \\
&&\left( V-V_{0}+E_{0}-E\right) \phi =0.  \notag
\end{eqnarray}
In the exactly solvable case, the potentials $V$ and $V_{0}$ are same. In the
quasi exactly solvable case, the functional form of the potential remain same
for all states, but a parameter changes from state to state. Let us
illustrate our method on the separable systems. In the case $f=0$ the
potentials can be written as
\begin{equation}
V-V_{0}=g(x)+h(y),  \label{eq:29}
\end{equation}
then substituting $\phi =R(x)Q(y)$ we obtain
\begin{eqnarray}
-\frac{d^{2}R}{dx^{2}}-2\frac{dW}{dx}\frac{dR}{dx}+(g(x)+E_{0}-E+c_{1})R &=&0
\notag \\
-\frac{d^{2}Q}{dy^{2}}-2\frac{dF}{dx}\frac{dQ}{dy}+(h(y)-c_{1})Q &=&0.
\label{eq:30}
\end{eqnarray}
In the following we illustrate our method on two examples.

\subsection{The model potential I}

Consider the following model potential with the parameter $\alpha $:
\begin{equation}
V(x,y;\alpha )=A^{2}x^{4}+ABx^{2}+C^{2}y^{2}+2A\alpha x.  \label{eq:31}
\end{equation}
The Schr\"{o}dinger equation with this potential is separable and can be
solved by reduction to two one-dimensional systems. The parameter $\alpha $
is depends on the state of the wave function. Let us call it the ``state
parameter''. The wave function of the ground state with the energy $E_{0}$
can be written in the form
\begin{eqnarray}
\psi _{0}(x,y) &=&\exp \left( -\frac{Ax^{3}}{3}-\frac{Cy^{2}}{2}-Bx\right)
\notag \\
E_{0} &=&C-B^{2}.  \label{eq:32}
\end{eqnarray}
The wave functions of the excited states with the energy $E$ we write as
\begin{equation}
\psi (x,y)=R(x)Q(y)\psi _{0}(x,y).  \label{eq:33}
\end{equation}
Substituting (\ref{eq:30}) in (\ref{eq:24}) with the potential (\ref{eq:31}%
), then after a straightforward calculation we obtain two one-dimensional
differential equations:
\begin{mathletters}
\begin{eqnarray}
-\frac{d^{2}Q}{dy^{2}}+2By\frac{dQ}{dy}-c_{1}Q &=&0  \label{eq:34a} \\
-\frac{d^{2}R}{dx^{2}}+2(C+Ax^{2})\frac{dR}{dx}+(E-E_{0}+c_{1}+2Ax(1-\alpha
))R &=&0,  \label{eq:34b}
\end{eqnarray}
where $c_{1}$ is a constant. The first equation is exactly solvable and its
solution is given by
\end{mathletters}
\begin{equation}
    Q(y)=N_{11}F_{1}\left( \frac{1}{2}-\frac{c_{1}}{4B},\frac{3}{2}%
    ,By^{2}\right) +N_{21}F_{1}\left( -\frac{c_{1}}{4B},\frac{1}{2}%
    ,By^{2}\right) ,  \label{eq:35}
\end{equation}
where $N_{1}$ and $N_{2}$ are normalization constants and $_{1}F_{1}\left(
\cdots \right) $ is a confluent hypergeometric function. The second equation
is a quasi-exactly solvable differential equation. The following
combinations of the operators of the Lie algebra
\begin{equation}
T=-J_{-}^{2}+2BJ_{-}+2AJ_{+}\quad \text{TR}=(E_{0}-E-c_{1})R  \label{eq:36}
\end{equation}
is equivalent with (\ref{eq:34b}) if $\alpha =j+1.$ Following the procedure
given in section 2, the spectrum of the system can be obtained by using the
following recurrence relation:
\begin{eqnarray}
&&-m(m-1)P_{m-2}(E)+2BmP_{m-1}(E)+  \label{eq:37} \\
&&2A(m-j)P_{m+1}(E)-(E_{0}-E-c_{1})P_{m}(E)=0.  \notag
\end{eqnarray}
Some of the polynomial is given by
\begin{eqnarray}
    P_{1}(E) &=&\left( c_{1}-E_{0}+E\right)   \notag \\
    P_{2}(E) &=&\left( c_{1}-E_{0}+E\right) ^{2}+4AB  \notag \\
    P_{3}(E) &=&\left( c_{1}-E_{0}+E\right) ^{3}+16AB\left( c_{1}-E_{0}+E\right)
    -16A^{2}  \notag \\
    P_{4}(E) &=&\left( c_{1}-E_{0}+E\right) ^{4}+40AB\left( c_{1}-E_{0}+E\right)
    ^{2}-  \label{eq:38} \\
    &&96A^{2}\left( c_{1}-E_{0}+E\right) +144A^{2}B^{2},  \notag
\end{eqnarray}
for $j=0,1,2,3$, respectively. The roots of the polynomials $P_{j}(E)$
produce the spectrum of the system.

\subsection{The model potential II}

The last example is the following double-well potential:
\begin{equation}
V(x,y;\alpha ;\beta
)=A_{1}^{2}x^{4}+A_{2}^{2}y^{4}+2A_{1}B_{1}x^{2}+2A_{2}B_{2}y^{2}-2A_{1}%
\alpha x-2A_{2}\beta y.  \label{eq:39}
\end{equation}
As we mentioned before, $\alpha $ and $\beta $ are state parameters. The
ground state wave function is given by
\begin{eqnarray}
    \psi _{0}(x,y) &=&\exp \left( -\frac{A_{1}x^{3}}{3}-\frac{A_{2}y^{3}}{3}%
    -B_{1}x-B_{2}y\right);   \notag \\
    E_{0} &=&B_{1}^{2}+B_{2}^{2}.  \label{eq:40}
\end{eqnarray}
Since the Schr\"{o}dinger equation is separable with the given potential,
the wave function can be written as
\begin{equation}
\psi (x,y)=R(x)\,Q(y)\,\psi _{0}.  \label{eq:41}
\end{equation}
Insertion of (\ref{eq:41}) into (\ref{eq:24}) with the potential (\ref{eq:39}%
) leads to the following two differential equations:
\begin{mathletters}
\begin{eqnarray}
-\frac{d^{2}R}{dx^{2}}+2\left( B_{1}+A_{1}x^{2}\right) \frac{dR}{dx}+\left(
E-E_{0}+c_{1}+2A_{1}(1-\alpha )\,x\right)  &=&0  \label{eq:42a} \\
-\frac{d^{2}Q}{dy^{2}}+2\left( B_{2}+A_{2}y^{2}\right) \frac{dQ}{dy}-\left(
c_{1}-2A_{2}(1-\beta )\,y\right)  &=&0.  \label{eq:42b}
\end{eqnarray}
These equations are QES.

\section{Conclusion}

The QES of \ the double-well potential in one and two dimensions have been
discussed. The solution of Schr\"{o}dinger equations which does not admit
separation of variables are left for treatment in future work.

\end{mathletters}

\end{document}